# One- and two-dimensional photo-imprinted diffraction gratings for manipulating terahertz waves

**Ioannis Chatzakis[1*], Philippe Tassin[1], Liang Luo[1], Nian-Hai Shen[1], Lei Zhang[1], Jigang Wang[1], Thomas Koschny[1], and Costas M. Soukoulis[1,2]**

[1] Ames Laboratory—U.S. DOE and Department of Physics and Astronomy, Iowa State University, Ames, IA 50011, USA

[2] Institute of Electronic Structure and Lasers (IESL), FORTH, 71110 Heraklion, Crete, Greece

**Emerging technology based on artificial materials containing metallic structures has raised the prospect for unprecedented control of terahertz waves through components like filters, absorbers and polarizers. The functionality of these devices is static by the very nature of their metallic or polaritonic composition, although some degree of tunability can be achieved by incorporating electrically biased semiconductors. Here, we demonstrate a photonic structure by projecting the optical image of a metal mask onto a thin GaAs substrate using a femtosecond pulsed laser source. We show that the resulting high-contrast pattern of photo-excited carriers can create diffractive elements operating in transmission. With the metal mask replaced by a digital micromirror device, our photo-imprinted photonic structures provide a route to terahertz components with reconfigurable functionality.**

---

[*] Present address: Department of Materials Science and Engineering, Stanford University, Stanford, California 94305, USA

The terahertz spectral range of the electromagnetic spectrum—loosely defined from about 100 GHz to 10 THz—has long been an inaccessible region in between the successful realms of electronics and photonics, because of the lack of efficient and compact sources and detectors for terahertz radiation. In the past few decades, however, the development of technologies like quantum-cascade lasers[1-3], terahertz wave generation through nonlinear crystals[4] and terahertz time-domain spectroscopy[5-6] has enabled the exploration of terahertz science and the rapid rise of terahertz imaging and spectroscopy for, amongst others, biomedical and security applications[7-8]. Controlling terahertz radiation has proven to be more difficult, although recent breakthroughs in waveguiding[9] and manipulating[10-15] terahertz waves have been reported. A major driving force behind these breakthroughs is the progress that scientists have made in artificial materials containing conductive structures, for example, frequency-selective surfaces[16], metamaterials[17-19] and photonic crystals[20-22]. By carefully tailoring the response of these artificial materials, it is possible to create terahertz devices like filters[10-11], absorbers[12-13] and polarizers[14-15]. It is even possible to achieve tunability by incorporating semiconductors with electrical biasing into the artificial materials[23-25] or the ability to switch between two different states[26-29], but these devices have essentially a specific functionality determined at the time of their design.

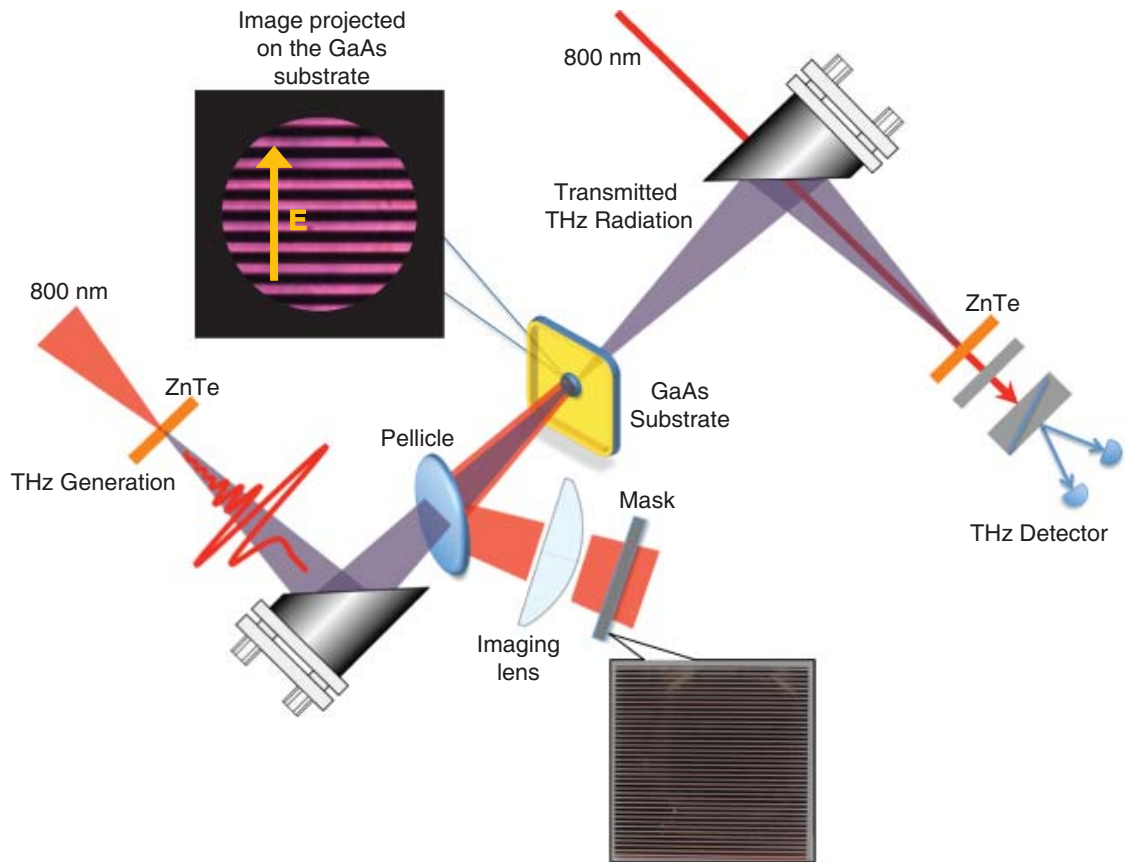

**Figure 1 | The experimental setup providing the basis for reconfigurable diffractive elements for terahertz waves. Optically projecting a mask on a GaAs substrate creates a grating in carrier concentration in the surface layer of the substrate.**

In this Letter, we develop an approach for a reconfigurable terahertz component, which can be switched between a wide class of functionalities, such as beam shaping, beam steering, and wavefront correction. The foundation of this approach, which is based on diffraction gratings, is the experiment outlined in Fig. 1. An 800 nm femtosecond-pulsed laser beam illuminates a metal mask and the image of the mask is projected with a lens onto a 1-mm-thin GaAs substrate. This creates a pattern of illuminated and dark regions on the substrate as shown in the inset of Fig. 1. Where illuminated, the pump beam creates carriers in the substrate by photo-excitation and the result is a photo-imprinted conductive pattern of free carriers inside the GaAs substrate, which we can use to manipulate terahertz waves. Being made from a copper film that is essentially impenetrable to the pump beam, the metal mask enables us in principle to create high-contrast light patterns on the GaAs substrate only limited by diffraction of the optical pump beam.

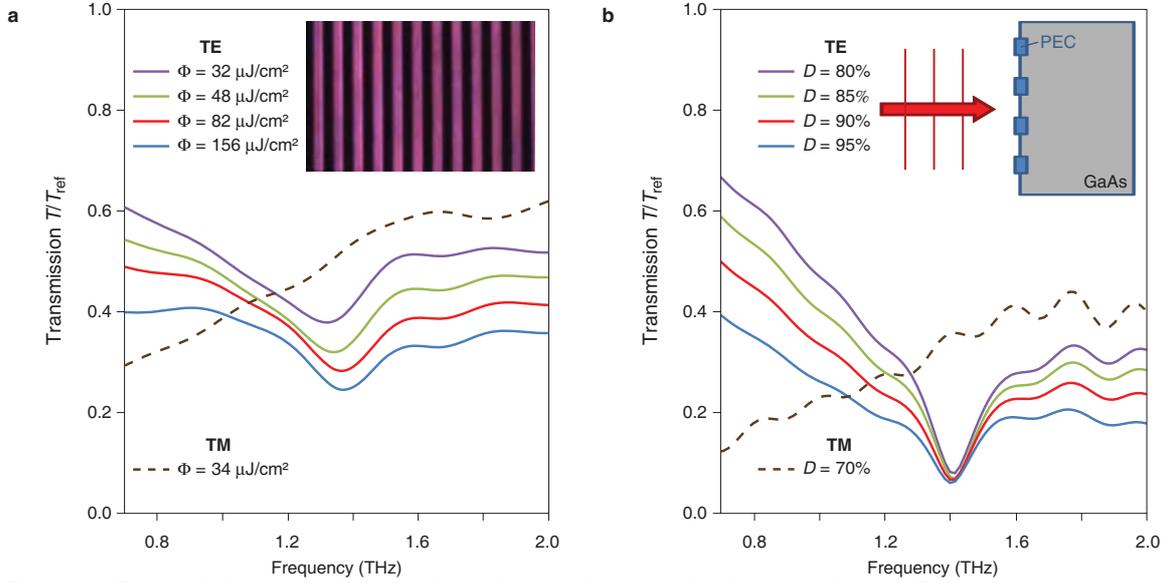

**Figure 2 | Terahertz transmission spectra of the photo-imprinted linear grating. a,** Experimental transmission spectra of a grating (lattice constant $a$ = 60 µm) with duty cycle $D$ = 50% for several pump fluencies. The inset shows the image of the metal mask at the position of the sample measured with a CCD camera. **b,** Simulation results from a simplified model of a diffraction grating (lattice constant $a$ = 60 µm) consisting of perfectly electrically conducting wires for different duty cycles $D$.

We subsequently show that the resulting photo-imprinted patterns can be used as diffractive elements for terahertz waves operating in transmission, which is highly desirable for many applications. Terahertz pulses, generated by optical rectification of the femtosecond laser pulses in a ZnTe crystal, are focused on the GaAs sample and the transmitted terahertz waves are detected by means of electro-optic sampling in a second ZnTe crystal. The terahertz pulse duration is approximately 2 ps, which is much smaller than the recombination time (≈ 1 ns) and the diffusion time (in 1 ns the carriers diffuse over a distance less than 2 µm) of the photocarriers, so the photo-imprinted grating can be considered to be essentially static for a single terahertz pulse. We start with a copper mask creating a photo-imprinted linear grating with lattice constant of 60 µm and a duty cycle of $D$ = 50%. From the terahertz transmission spectra for a set of different pump fluences, shown in Fig. 2a, we observe a pronounced spectral feature with strongly reduced transmission at about $f$ = 1.39 THz when the electric field is polarized perpendicular to the grating lines (TE). For the other polarization (TM), no sharp spectral feature is observed. We also observe higher background transmission levels for low pump fluence—i.e., when few photocarriers are created—and low background transmission for higher pump fluence.

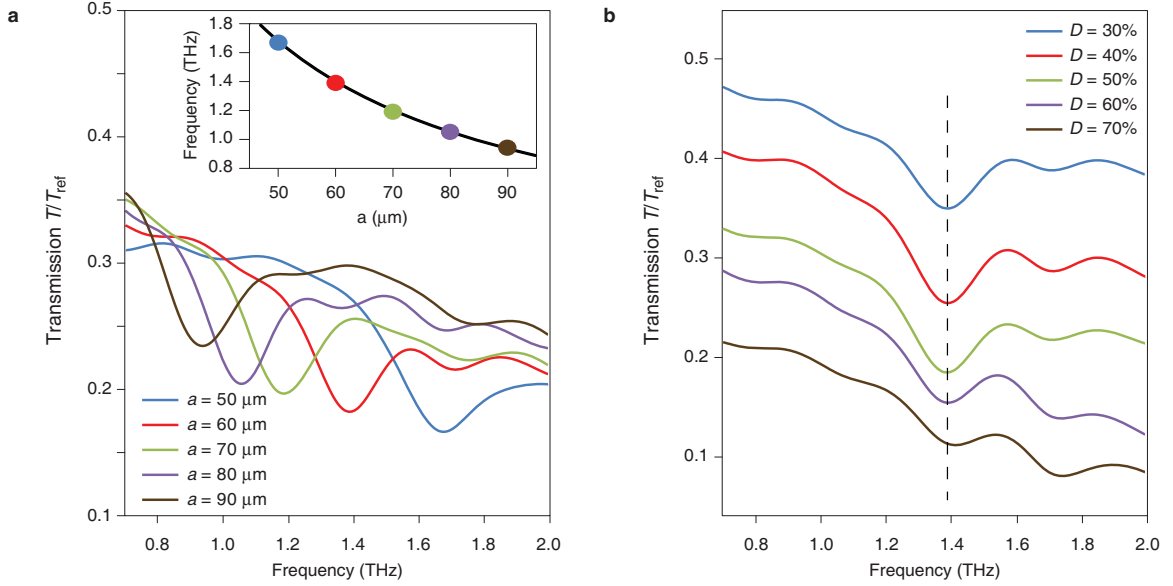

**Figure 3 | Is the spectral feature due to coupling to the first diffraction order or due to a quasistatic electric dipole resonance? a,** Experimental transmission spectra for masks with varying lattice constant (Fluence is Φ = 96 µJ/cm²; duty cycle is $D$ = 50%). The inset shows that the transmission dip frequencies are inversely proportional to the lattice constant (consistent with diffraction). The black line is the cut-off frequency of the first diffraction order in GaAs, $f = c/(n_{GaAs} \cdot a)$ **b,** Experimental transmission spectra for masks with varying duty cycle (Φ = 96 µJ/cm²; $a$ = 60 µm). The spectral position of the dips does not depend on the duty cycle (consistent with diffraction; inconsistent with a cut-wire resonance).

The two most plausible mechanisms possibly behind the spectral feature are (i) coupling to the first diffraction order in GaAs and (ii) a quasistatic electric dipole resonance in the cut wire elements created by the neighbouring stripes of the grating. In order to distinguish between both mechanisms, we have fabricated a set of masks with varying lattice constant (duty cycle fixed at $D$ = 50%) and a set with varying duty cycle (lattice constant fixed at $a$ = 60 µm). The transmission spectra for the set with varying lattice constant, displayed in Fig. 3a, show a redshift of the spectral feature for increasing lattice constant. The spectral positions are inversely proportional to the lattice constants (see inset of Fig. 3a) and consistent with the cut-off frequency of the first diffraction order in GaAs, $f = c/(n_{GaAs} \cdot a)$, with $n_{GaAs}$ = 3.56 for GaAs at 1 THz. The transmission spectra for the set with varying duty cycle, displayed in Fig. 3b, show there is no shift in spectral position when the distance between the wires is changed. This observation rules out the electric dipole resonance, which has a resonance frequency that strongly depends on the capacitance between the cut wires[30]. The redshift of the spectral minimum when the grating period is increased points to application of our structure for reconfigurable terahertz filters.

Further evidence that our photo-imprinted gratings are purely diffraction-based and do not suffer from additional complications like quasistatic resonances or surface modes comes from simulation results involving a simple linear grating of perfectly conducting wires on a GaAs substrate (see Fig. 2b). The resulting transmission spectra agree qualitatively with the experimental spectra (Fig. 2a)—both in the spectral position of the dip and the overall shape of the transmission spectra. No surface modes or quasistatic resonances are seen in the electric field distribution obtained from the computer simulations. In addition, comparison of the experimental results of Fig. 2a and the simple model in Fig. 2b tells us that different pump fluence values result in diffraction gratings with different diffraction efficiency; i.e., the pump fluence provides us with a straightforward control over the properties of the diffractive element.

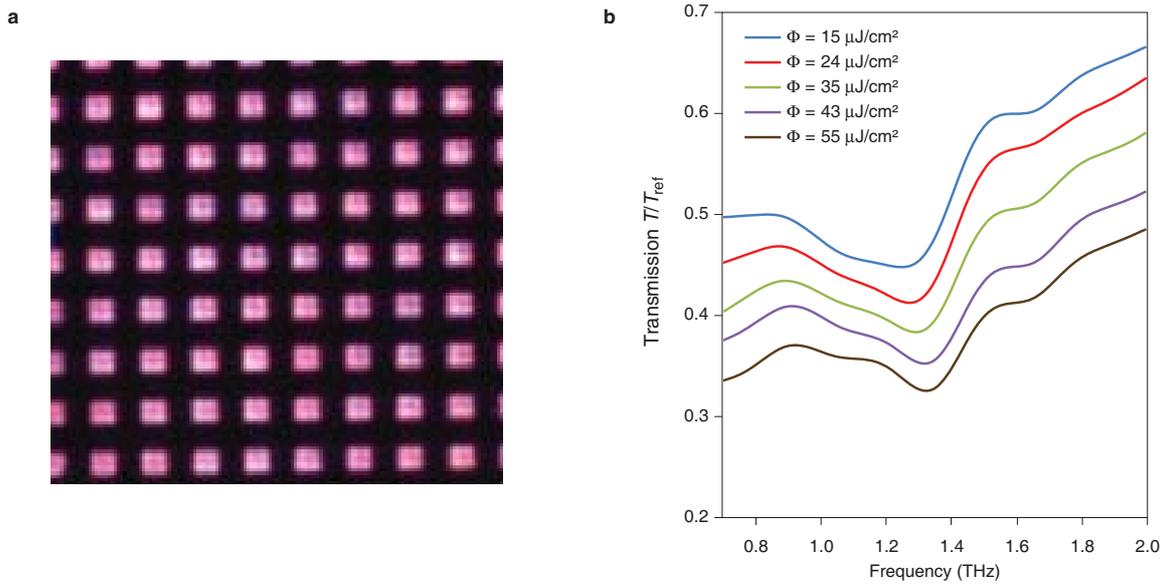

**Figure 4 | A two-dimensional photo-imprinted diffractive element. a,** Optical image of the metal mask at the position of the sample measured with a CCD camera. **b,** Terahertz transmission spectra showing the transmission dip at the diffraction edge due to coupling into the first diffraction order.

Not only can we create photo-imprinted linear gratings, but we can also generate two-dimensional diffractive elements on the GaAs substrate. One example—a square-lattice pattern with islands of photocarriers—is shown in Fig. 4a (optical image of the mask obtained with a CCD camera at the sample position). The resulting transmission spectra again show a clear minimum at the diffraction edge associated with coupling of energy into the first diffraction order. In fact we can create arbitrary patterns and these experiments therefore open the door to designing diffractive elements with well-established techniques from Fourier optics, which allow determining the required amplitude gratings for fairly general functions including beam shaping (Fresnel lenses, spatial filtering, wavefront correction) and beam steering, fan-outs, etc.

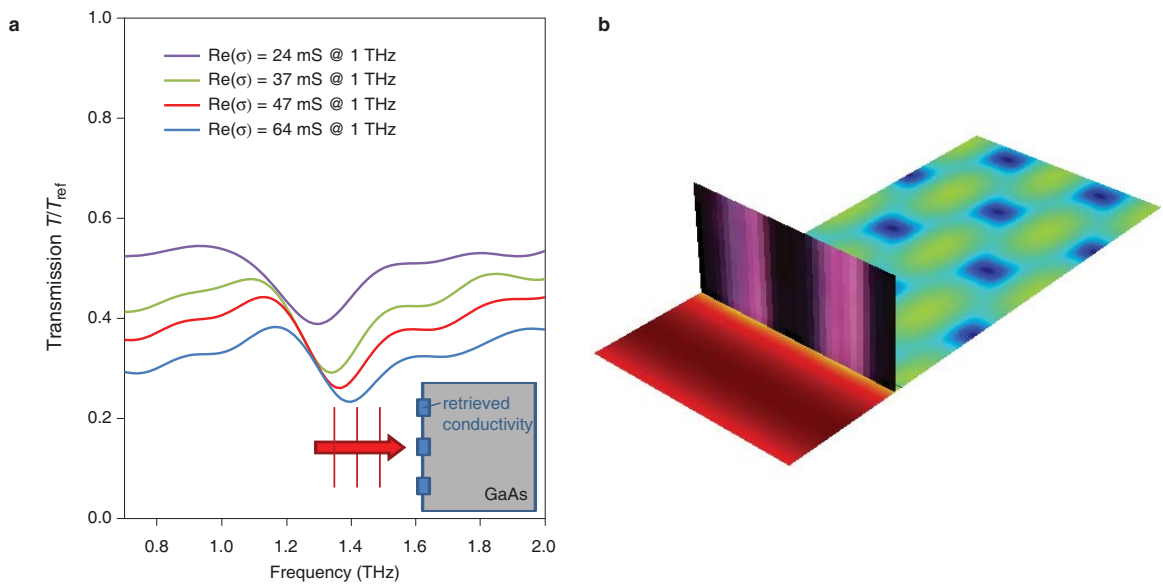

**Figure 5 | Results from computer simulations of the photo-imprinted carrier grating in the GaAs substrate taking into account the lossy nature of the photo-excited carriers. a,** Transmission spectra, in excellent agreement with the experimental results. **b,** Electric field component perpendicular to the grating at $f$ = 1.6 THz. We recognize the checkerboard pattern typical for the interference between the zeroth- and first-order diffracted waves.

Finally, we have also performed realistic computer simulations of the experimental setup (the linear grating) that take into account the lossy nature of the photo-excited carriers. We now model the photo-excited free carriers as a conductive sheet with a frequency-dependent conductivity determined from measurements of illuminated GaAs substrates without any mask in place. Fig. 5a plots the transmission spectra from these realistic simulations for several values of the carrier density—they are in excellent agreement with the experiments (Fig. 2a). We have accounted for the finite contrast ratio in the image of the mask. For high pump fluences, this will result in the creation of some photocarriers in the dark regions and, thus, in diffraction gratings with finite amplitude contrast. As a consequence, we get smaller dip amplitude than in the idealized simulation of Fig. 2b. In addition, the background transmission now depends characteristically on the carrier density (pump fluence) and the transmission dip is slightly blueshifted for higher carrier density, as observed in the experiment. Finally, Fig. 5b plots the electric field inside the GaAs substrate. We observe the typical checkerboard pattern created by the interference of the transmitted and diffracted beams. We do not find any resonant fields in the structure, thus avoiding the dissipative loss that that would be produced by surface waves and resonant metamaterials[31]. The simulations described in this paragraph will also be crucial in the design of the more intricate diffractive elements mentioned above.

The demonstrated photo-imprinted diffractive elements can be made reconfigurable by replacing the fixed metal mask employed here by a device with spatially controlled transmission for the optical pump beam. One possibility is a liquid-crystal spatial light modulator as recently utilized by Okada and Tanaka[29] for tunable terahertz devices, but liquid crystals tend to have limited contrast, degrading the quality of the diffraction gratings as shown above. In order to retain the excellent contrast ratios provided by the basically impenetrable copper mask, we could replace the mask by a digital micromirror device (DMD), a technology that is now widely used in DLP projection equipment. The reconfiguration rate that can be achieved in this way will be limited by the DMD mirror switching time of about 20 μs. The previous grating must also be erased, of course, but with a typical recombination time of the photocarriers of 1 nanosecond this will not be a limiting factor. This allows for a reconfiguration rate of the terahertz device of 50 kHz, sufficient for both sensing and imaging applications relying on terahertz science.

**Methods**

**Terahertz measurement setup** We refer to Fig. 1 for an outline of the experimental setup. The metal mask was projected onto a 520-μm-thick high-resistivity GaAs substrate using an optical imaging system with a magnification factor of 1/8—constructed from an aspherical lens with a focal length of 150mm. The illumination source for the projection was an 800 nm pulsed laser beam derived from a 1 kHz Ti:Sapphire amplifier (see refs. [32-33] for more details about the pump laser and the terahertz pulse generation), which was expanded to uniformly illuminate the mask. A CCD camera was used in the position of the GaAs wafer to determine the conjugate plane of the mask (and providing us with the pictures in Figs. 2 and 4).

A small fraction of the 800 nm pump beam with approximately 40 fs pulse duration was used to generate and detect terahertz pulses. The phase-locked THz field transients, generated via optical rectification and detected by electro-optic sampling in 1-mm-thick ZnTe crystals, were used as a probe to determine the transmission of the sample. The transmitted pulses were recorded and cut after 6.5 ps to avoid multiple reflections in the GaAs substrate. The resulting signals were zero-padded and Fourier transformed, resulting in a frequency resolution of 0.15 THz. Finally, this experiment was performed for the illuminated sample with mask and for the bare GaAs substrate (unilluminated, no mask), and the ratio of both spectra is shown in Figs. 2, 3, and 4.

**Mask fabrication** The metal mask was made by sputtering a 2-μm-thick copper film on a 1.1-mm-thick borosilicate glass substrate and the diffraction gratings—8 times enlarged with respect to the photo-imprinted gratings—were defined with photolithography and chemically etched.

**Ideal diffraction grating model** The grating with duty cycle $D$ was modelled by a thin strip of PEC with width $w = D*60$ μm on a GaAs substrate ($n = 3.56$). Periodic boundary conditions were applied to represent the translational symmetry of the grating. Port boundary conditions were used to terminate the simulation domain in the longitudinal directions, with the receiving port matched to the modes in the GaAs substrate. The transmission spectrum was obtained using a finite-element electromagnetics solver (CST Microwave Studio).

**Realistic numerical simulations** The grating with 50% duty cycle was modelled by a conductive sheet. The frequency-dependent conductivity was derived from transmission experiments of illuminated substrates (without the mask in place). In the "unilluminated" regions a conductive sheet with 10 times smaller conductivity was applied to account for the finite contrast ratio in the experiment. Periodic boundary conditions were applied to represent the translational symmetry of the grating and port boundary conditions were used to terminate the simulation domain in the longitudinal directions, with the receiving port matched to the modes in the GaAs substrate. The transmission spectrum was obtained using finite-element electromagnetics solvers (COMSOL and CST). A pulse of the form $\sin(6.4\ \text{THz}*t + (1.6\ \text{THz}*t)^2) \exp(-(1.6\ \text{THz}*t)^2)$ was subsequently transmitted through a system with the simulated spectra, and the transmitted pulses were cut, zero-padded, and Fourier transformed as in the experiments. The reference was obtained from a simulation of a bare GaAs substrate with index of refraction of 3.56 (also retrieved from experiments), and the ratio of the transmission spectra is shown in the figures.

**Acknowledgments**

Work at Ames Laboratory was partially supported by the U.S. Department of Energy, Office of Basic Energy Science, Division of Materials Sciences and Engineering (Ames Laboratory is operated for the U.S. Department of Energy by Iowa State University under Contract No. DE-AC02-07CH11358) (experiments) and by the U.S. Office of Naval Research, Award No. N00014-10-1-0925 (theory).